\documentclass[numreferences]{kluwer}

\newdisplay{guess}{Conjecture}

\begin{document}

\begin{article}

\begin{opening}

\title{Extended discrete KP hierarchy and its dispersionless limit\thanks{This
work was supported by INTAS grant 2000-15 and RFBR grant
03-01-00102.}}

\author{ANDREI K. \surname{SVININ}}

\runningauthor{ANDREI K. SVININ}

\runningtitle{Extended discrete KP hierarchy and its dispersionless limit}

\institute{Institute of System Dynamics and Control Theory, Siberian
Branch of Russian Academy of Sciences, P.O. Box 1233, 664033 Irkutsk,
Russia}

\date{}

\begin{abstract}
We exhibit the dispersionless limit of the extended discrete KP
hierarchy.
\end{abstract}


\end{opening}

\section{Introduction}

There are many activities concerning with dispersionless hierarchies.
The interest to these hierarchies comes, in particular, from string
theory and related areas (see, for example \cite{takasaki1},
\cite{kharchev}, \cite{boyarsky}). Of course, one can also mention
many other interesting themes like the study of the system of
hydrodynamic type \cite{kodama1}, the theory of conformal maps
\cite{wiegmann}, \cite{mineev} etc.

This letter is designed to present the extension of the
dispersionless discrete KP hierarchy in a hope that this can be also
useful in modern string physics or elsewhere. To this aim, we
formulate the extended discrete KP (edKP) hierarchy in a suitable
form in Section 2. In our opinion this representation is well
background for performing dispersionless limit and for deriving  Lax
equations. This is doing in Section 3. In Appendix we display some
evolution equations coded in corresponding Lax equations.

\section{edKP hierarchy}

We consider the space of analytic functions $\{f(s) :
s\in\mathbb{R}\}$ whose domain of definition is restricted to
$\mathbb{Z}$. We write $f=f(i)\in R$. On $R$ the shift operator
$\Lambda = \exp(\partial/\partial s)$ acts as $(\Lambda f)(i) =
f(i+1)$.

Let us consider the space of pseudo-difference ($\vartriangle$DO)
operators $\mathcal{D}=R[\Lambda,\Lambda^{-1})$ of the form
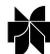
\[
P = \sum_{j=-\infty}^Np_j\Lambda^j,\;\; p_j\in R
\]
where $N\in\mathbb{Z}$ is treated as an order of $P$.

Phase-space of the discrete KP hierarchy can be identified with that
of Zakharov-Shabat discrete dressing operators
\[
\mathcal{M} = \{S = I + \sum_{k\geq 1}w_k\Lambda^{-k} : w_k\in R\}.
\]

One considers the splitting $\mathcal{D}
=\mathcal{D}_{+}^{(1)}\oplus\mathcal{D}_{-}^{(1)}$ with $
\mathcal{D}_{+}^{(1)}\equiv<I, \Lambda, \Lambda^2,...>$  and $
\mathcal{D}_{-}^{(1)}\equiv<\Lambda^{-1}, \Lambda^{-2},...>$.  The
discrete KP hierarchy is given by Sato-Wilson
\[
\frac{\partial S}{\partial t_p^{(1)}} = -\pi_{-}^{(1)}(Q^p)S =
\pi_{+}^{(1)}(Q^p)S - S\Lambda^p
\]
or equivalently by Lax equations
\[
\frac{\partial Q}{\partial t_p^{(1)}} = [\pi_{+}^{(1)}(Q^p), Q] = [Q,
\pi_{-}^{(1)}(Q^p)]
\]
where the symbols $\pi_{+}^{(1)}$ and $\pi_{-}^{(1)}$ stand for
projections of any $\vartriangle$DO on the space
$\mathcal{D}_{+}^{(1)}$ and $\mathcal{D}_{-}^{(1)}$, respectively.
The discrete Lax operator is defined as usual:
\[
Q=S\Lambda S^{-1} = \Lambda + \sum_{k\geq 1}q_k\Lambda^{1-k}.
\]

Fixing any point $\mathcal{M}$ and integer $n\geq 2$, we consider the
splitting
$\mathcal{D}=\mathcal{D}_{+}^{(n)}\oplus\mathcal{D}_{-}^{(n)}$ with
\[
\mathcal{D}_{+}^{(n)}\equiv<I, \Lambda^nQ^{1-n},
\Lambda^{2n}Q^{2(1-n)},...>
\]
and
\[
\mathcal{D}_{-}^{(n)}\equiv<\Lambda^{-n}Q^{n-1},
\Lambda^{-2n}Q^{2(n-1)},...>.
\]
One can see that this splitting (for $n\geq 2$) essentially depends
on the point of the phase-space $\mathcal{M}$. For any $P\in
\mathcal{D}$ one can write
\[
P=p_N\Lambda^{Nn}Q^{N(1-n)}
+p^{'}_{N-1}\Lambda^{(N-1)n}Q^{(N-1)(1-n)} + ...
\]
This representation is correctly defined since
$\Lambda^{kn}Q^{k(1-n)}$ is a $\vartriangle$DO of an order $k$ and
therefore one can step-by-step calculate the coefficients
$p^{'}_{N-1}, p^{'}_{N-2},...$ as polynomials of $p_k$ and $q_k$.

We define edKP hierarchy by
\[
\frac{\partial S}{\partial t_p^{(n)}} = -\pi_{-}^{(n)}(Q^p)S =
\pi_{+}^{(n)}(Q^p)S - S\Lambda^p
\]
or equivalently by
\begin{equation}
\frac{\partial Q}{\partial t_p^{(n)}} = [\pi_{+}^{(n)}(Q^p), Q] = [Q,
\pi_{-}^{(n)}(Q^p)]
\label{Lax}
\end{equation}
with $\pi_{+}^{(1)}$
and $\pi_{-}^{(1)}$ being projections on corresponding spaces. One
can easy prove that these equations are correctly defined making of
use standard reasonings.

Let $\chi(z) \equiv (z^i)_{i\in\mathbb{Z}}$ where $z$ can be
interpreted as a (formal) spectral parameter. Define $w(z) =
S\chi(z)$ and formal Baker-Akhiezer function $\Psi(t, z) =
w(z)e^{\xi(t,z)}$, where $\xi(t,
z)\equiv\sum_{n,p=1}^{\infty}t_p^{(n)}z^p$.

It is common of knowledge that Lax equations like (\ref{Lax}) can be
understood as consistency condition of the linear system
\[
Q\Psi=z\Psi,\;\; \frac{\partial \Psi}{\partial t_p^{(n)}}
=\pi_{+}^{(n)}(Q^p)\Psi.
\]

In what follows we are going to show that this system can be cast in
the form
\begin{equation}
Q_{(n)}\Psi=z\Psi,\;\; z^{p(n-1)}\frac{\partial \Psi}{\partial
t_p^{(n)}} =(Q^{pn}_{(n)})_{+}\Psi \label{12}
\end{equation}
with $Q_{(n)}\equiv S_{(n)}\Lambda S_{(n)}^{-1}$ where $S = I +
\sum_{k\geq 1}z^{k(n-1)}w_k\Lambda^{-kn}$. The symbol $+$ denotes
taking only nonnegative powers of $\Lambda$.

Multiplication $\chi(z)$ by $z$ is equivalent to the action of the
shift operator. Therefore one can easy to check that as a result of
action $S_{(n)}$ to $\chi(z)$ for any integer $n\geq 1$ is $w(z)$ and
from this we obtain that $Q_{(n)}w(z) = zw(z)$ or $Q_{(n)}\Psi =
z\Psi$. To succeed it is useful to define, on $\mathcal{M}$, the
functions $q_k^{(n,r)}$ through the relation
\[
Q^r_{(n)} = S_{(n)}\Lambda^r S_{(n)}^{-1} = \Lambda^r + \sum_{k\geq
1}z^{k(n-1)}q_k^{(n,r)}\Lambda^{r-kn}.
\]
By definition, we have $Q^r_{(n)}\Psi = z^r\Psi$.

Let us prove that for an arbitrary pair of integers $n, n^{'}\geq 1$
the relation
\begin{equation}
Q^r_{(n^{'})} =  \Lambda^r + \sum_{k\geq
1}z^{k(n^{'}-1)}q_k^{(n,r)}\Lambda^{r-kn}Q^{k(n-n^{'})}_{(n^{'})}
\label{RHSLHS}
\end{equation}
is valid.

Firstly we have
\begin{eqnarray}
Q^r_{(n^{'})}\Psi&=&\left\{\Lambda^r + \sum_{k\geq
1}z^{k(n^{'}-1)}q_k^{(n,r)}\Lambda^{r-kn}Q^{k(n-n^{'})}_{(n^{'})}\right\}\Psi
\nonumber \\[0.3cm]
&=&\left\{\Lambda^r + \sum_{k\geq
1}z^{k(n-1)}q_k^{(n,r)}\Lambda^{r-kn}\right\}\Psi \nonumber \\[0.3cm]
&=&Q_{(n)}^r\Psi=z^r\Psi. \nonumber
\end{eqnarray}
Moreover it is obvious that LHS and RHS of (\ref{RHSLHS}) are of the
same form.

More explicitly one can rewrite (\ref{RHSLHS}) as
\[
q_k^{(n^{'},r)}(i) = q_k^{(n,r)}(i) +
\sum_{m=1}^{k-1}q_m^{(n,r)}(i)q_{k-m}^{(n^{'},m(n-n^{'}))}(i+r-mn).
\]
In particular $q_1^{(n^{'},r)}(i) = q_1^{(n,r)}(i)$. Putting in
(\ref{RHSLHS}) $n^{'}=1$, we obtain
\[
Q^r =  \Lambda^r + \sum_{k\geq 1}q_k^{(n,r)}\Lambda^{r-kn}Q^{k(n-1)}.
\]
Let $r=pn$ with some integer $p\geq 1$, then
\[
Q^{pn} =  \Lambda^{pn} + \sum_{k\geq
1}q_k^{(n,pn)}\Lambda^{(p-k)n}Q^{k(n-1)}.
\]
Multiplying LHS and RHS of the latter by $Q^{p(1-n)}$, we get
\[
Q^{pn}\cdot Q^{p(1-n)} = Q^p = \Lambda^{pn}Q^{p(1-n)} + \sum_{k\geq
1}q_k^{(n,pn)}\Lambda^{(p-k)n}Q^{(p-k)(1-n)}
\]
The latter formula gives suitable form to split $Q^p$ on `positive'
and `negative' parts. In particular
\[
\pi_{+}^{(n)}(Q^p) = \Lambda^{pn}Q^{p(1-n)} + \sum_{k=1}^p
q_k^{(n,pn)}\Lambda^{(p-k)n}Q^{(p-k)(1-n)}.
\]
From this we have
\[
\pi_{+}^{(n)}(Q^p)\Psi = z^{p(1-n)}\left\{ \Lambda^{pn} +
\sum_{k=1}^pz^{k(n-1)}q_k^{(n,pn)}\Lambda^{(p-k)n}\right\}\Psi
\]
\[
\equiv z^{p(1-n)}(Q_{(n)}^{pn})_{+}\Psi.
\]
The latter gives the second equation in (\ref{12}). The consistency
condition of (\ref{12}) is expressed in the form of the following Lax
equations:
\begin{equation}
z^{p(n-1)}\frac{\partial Q_{(n)}}{\partial t_p^{(n)}} =
[(Q^{pn}_{(n)})_{+}, Q_{(n)}]. \label{Leq}
\end{equation}

Let us spend some lines to give bibliographical remarks concerning
(\ref{Leq}). These equations, in fact, are equivalent to
Kupershmidt's gap KP hierarchy (see, for example \cite{ku}). More
exactly, he considers Lax operator with anti-normal ordering
\[
L = \Lambda + \Lambda^{1-\Gamma}\circ q_0 + \Lambda^{1-2\Gamma}\circ
q_1 + ...,\;\; \Gamma\geq 1
\]
and corresponding Lax equations
\begin{equation}
\frac{\partial L}{\partial t_p} = [L^{p\Gamma}_{+}, L]. \label{ku}
\end{equation}
In Ref. \cite{ku} the problem of integrable discretization of the
flows given by (\ref{ku}) is solved.

 In Refs. \cite{svinin1} and \cite{svinin2} we exhibit
two-parametric class of invariant submanifolds $\mathcal{S}_l^n$ of
the Darboux-KP chain \cite{mpz}. Moreover, we showed that double
intersections $\mathcal{S}_{n,r,l} =
\mathcal{S}_0^n\cap\mathcal{S}_{l-1}^{ln-r}$ lead to a broad class of
integrable lattices over finite number of fields (see also
\cite{svinin3}, \cite{svinin4} and references therein ). All these
systems share Lax representation of the form
\[
z^{p(n-1)}\frac{\partial Q_{(n)}^r}{\partial t_p^{(n)}} =
[(Q^{pn}_{(n)})_{+}, Q_{(n)}^r].
\]
with restricted Lax operator
\[
Q^r_{(n)} =  \Lambda^{r} + \sum_{k=1}^l z^{k(n-1)}
q_k^{(n,r)}\Lambda^{r-kn}.
\]
It is important to note that $r$ in this case cannot be treated as a
power of $Q_{(n)}$.

The equations (\ref{Lax}) have its advantage that they clearly show
that the flows labeled by integers $n\geq 2$ are defined on the same
phase-space as for the customary discrete KP hierarchy and in our
opinion this representation is more convenient for performing
dispersionless limit.

\section{Dispersionless edKP hierarchy}

Dispersionless limit of the discrete KP hierarchy or more generally
of the Toda lattice one \cite{kodama}, \cite{takasaki} is performed
by replacing
\[
\frac{\partial}{\partial t_p}\rightarrow\hbar\frac{\partial}{\partial
t_p},\;\; e^{\partial_s}\rightarrow e^{\hbar\partial_s},\;\;
q_k\rightarrow \hbar^{-1}q_k
\]
and tending $\hbar$ to zero. Then $\partial_s$ is replaced by formal
parameter $p$\footnote{Note that this $p$ has nothing to do with
integer $p$ labeled evolution parameters} and spectral problem turn
into
\[
z=e^p + q_1 + q_2e^{-p} + q_3e^{-2p} + ...
\]
In turn, the dispersionless discrete KP hierarchy is defined by
corresponding Lax equations
\[
\frac{\partial z}{\partial t_p^{(1)}} = \{\pi_{+}^{(1)}(z^p), z\} =
\{z, \pi_{-}^{(1)}(z^p)\}.
\]
The commutator $[\cdot, \cdot]$ on $\mathcal{D}$ in the
dispersionless limit is replaced by Poisson bracket
\[
\{f, g\} = \frac{\partial f}{\partial p}\frac{\partial g}{\partial s}
-\frac{\partial f}{\partial s}\frac{\partial g}{\partial p} =
k\left(\frac{\partial f}{\partial k}\frac{\partial g}{\partial s}
-\frac{\partial f}{\partial s}\frac{\partial g}{\partial
k}\right),\;\; k=e^p.
\]
Symbols $\pi_{+}^{(1)}$ and $\pi_{-}^{(1)}$ in this case stand for
projections on the spaces $ \tilde{\mathcal{D}}_{+}^{(1)}\equiv<1, k,
k^2,...>$ and $ \tilde{\mathcal{D}}_{-}^{(1)}\equiv<k^{-1},
k^{-2},...>$ which split on `positive' and `negative' parts the space
$\tilde{\mathcal{D}}$ of formal Laurent series of the form
\[
\tilde{P} = \sum_{j=-\infty}^N\tilde{p}_jk^j
\]
with some coefficients $\tilde{p}_k$ being analytic functions of the
variable $s$.

We can immediately to write down the equations which are a result of
dispersionless limit applied to edKP hierarchy in the form of
(\ref{Lax}). We have
\begin{equation}
\frac{\partial z}{\partial t_p^{(n)}} = \{\pi_{+}^{(n)}(z^p), z\} =
\{z, \pi_{-}^{(n)}(z^p)]. \label{dedKP}
\end{equation}
Now the symbols $\pi_{+}^{(n)}$ and $\pi_{-}^{(n)}$ are used to
denote projections on the spaces
\[
\tilde{\mathcal{D}}_{+}^{(n)}\equiv<1, z^{1-n}k^n,
z^{2(1-n)}k^{2n},...>
\]
and
\[
\tilde{\mathcal{D}}_{-}^{(n)}\equiv<z^{n-1}k^{-n},
z^{2(n-1)}k^{-2n},...>,
\]
respectively.

We aware that the (dispersionless) discrete KP hierarchy is only a
part of the (dispersionless) Toda one and we are going in future to
discuss extended version of the Toda lattice hierarchy.

\appendix

a. Let us give here explicit form of some functions $q_k^{(n,r)}$. We
calculate first three of them to wit
\[
q_1^{(n,r)}(i) = w_1(i) - w_1(i+r) = q_k^{(r)}(i),
\]
\[
q_2^{(n,r)}(i) = w_2(i) - w_2(i+r) + w_1(i+r-n)(w_1(i+r) - w_1(i))
\]
\[
=q_2^{(r)}(i) - q_1^{(r)}(i)q_1^{(n-1)}(i+r-n),
\]
\[
q_3^{(n,r)}(i) = w_3(i) - w_3(i+r) + w_1(i+r-2n)(w_2(i+r) - w_2(i))
\]
\[
+ w_2(i+r-n)(w_1(i+r) - w_1(i))
\]
\[
+w_1(i+r-2n)w_1(i+r-n)(w_1(i) - w_1(i+r))
\]
\[
=q_3^{(r)}(i) - q_1^{(r)}(i)q_2^{(n-1)}(i+r-n) -
q_2^{(r)}(i)q_1^{(2(n-1))}(i+r-2n)
\]
\[
+ q_1^{(r)}(i)q_1^{(n-1)}(i+r-n)q_1^{(2(n-1))}(i+r-2n).
\]
Here $q_k^{(r)}$'s are the coefficients of $r$-th power of $Q$, i.e.
\[
Q^r = \Lambda^r + q_1^{(r)} + q_2^{(r)}\Lambda^{-1} + ...
\]

b. Let us exhibit some examples of equations of motion coded in
(\ref{dedKP}) for $n=2$. We have the following:
\[
\frac{\partial q_1}{\partial t_1^{(2)}} =
2(q_2^{\prime}-q_1q_1^{\prime}),\;\; \frac{\partial q_2}{\partial
t_1^{(2)}} = 2(q_3^{\prime}-q_1q_2^{\prime} + q_1^2q_1^{\prime}),
\]
\[
\frac{\partial q_3}{\partial t_1^{(2)}} =
2(q_4^{\prime}-q_1q_3^{\prime} + (q_1^2 - q_2)q_2^{\prime} - (q_1^3 -
2q_1q_2-q_3)q_1^{\prime}),...
\]
\[
\frac{\partial q_1}{\partial t_2^{(2)}} = 4q_3^{\prime},\;\;
\frac{\partial q_2}{\partial t_2^{(2)}} = 4(q_4^{\prime} + (q_1^2 -
q_2)q_2^{\prime} - (q_1^3 - 2q_1q_2-q_3)q_1^{\prime}),...
\]

\end{article}

\end{document}